\documentclass[prd,10pt,aps,twocolumn,superscriptaddress,floatfix,notitlepage,nofootinbib,amssymb,amsmath]{revtex4-1}
\usepackage{mathtools}
\usepackage{epsfig}

\newcommand{\tE}{{\widetilde E}}

\newcommand{\beqn}{\begin{eqnarray}}
\newcommand{\eeqn}{\end{eqnarray}}
\newcommand{\eq}[1]{(\ref{#1})}
\newcommand{\jj}{{\mathrm{j}}}

\newcommand{\cL}{{\cal L}}

\newcommand{\bp}{\mathsf{p}}
\newcommand{\bE}{\mathsf{E}}

\newcommand{\lab}{{\mathrm {lab}}}

\newcommand{\Z}{{\mathbb Z}}
\newcommand{\R}{{\mathbb R}}
\newcommand{\N}{{\mathbb N}}

\newcommand{\bs}{\boldsymbol}

\newcommand{\avr}[1]{{\left\langle #1 \right\rangle}}

\def\bbbone{{\mathchoice {\rm 1\mskip-4mu l} {\rm 1\mskip-4mu l} {\rm 1\mskip-4.5mu l} {\rm 1\mskip-5mu l}}}

\begin{document}

\title{Effects of rotation and boundaries on chiral symmetry breaking of relativistic fermions}

\author{M. N. Chernodub}
\affiliation{Laboratoire de Math\'ematiques et Physique Th\'eorique UMR 7350, Universit\'e de Tours, 37200 France}
\affiliation{Laboratory of Physics of Living Matter, Far Eastern Federal University, Sukhanova 8, Vladivostok, 690950, Russia}
\author{Shinya Gongyo}
\affiliation{Laboratoire de Math\'ematiques et Physique Th\'eorique UMR 7350, Universit\'e de Tours, 37200 France}
\affiliation{Theoretical Research Division, Nishina Center, RIKEN, Saitama, Japan}

\begin{abstract}
In order to avoid unphysical causality-violating effects any rigidly rotating system must be bounded in directions transverse to the axis of rotation. We demonstrate that this requirement implies substantial dependence of properties of relativistically rotating system on the boundary conditions. We consider a system of interacting fermions described by the Nambu--Jona-Lasinio model in a space bounded by cylindrical surface of finite radius. In order to confine the fermions inside the cylinder we impose ``chiral'' MIT boundary conditions on its surface. These boundary conditions are parameterized by a continuous chiral angle $\Theta$. We find that at any value of $\Theta$ the chiral restoration temperature $T_c$ decreases as a quadratic function of the angular frequency $\Omega$. However, the position and the slope of the critical curve $T_c = T_c(\Omega)$ in the phase diagram depends noticeably on the value of the chiral angle.
\end{abstract}

\maketitle

\section{Introduction}

The phase structure of rotating systems plays an important role in large variety of physical environments, from trapped nonrelativistic bosonic cold atoms~\cite{Stringari:1996zz} to rapidly rotating neutron stars~\cite{Cook:1993qr} and noncentral heavy-ion collisions. The collisions may produce relativistically rotating quark-gluon plasma with large angular momentum~\cite{ref:HIC}. In addition, rotating fermionic matter possesses exotic anomalous transport phenomena~\cite{ref:CVE} which appear in astrophysical context~\cite{ref:Vilenkin} and in solid state physics~\cite{ref:Weyl}. 

Properties of rotating fermionic systems have been discussed both for free particles~\cite{Iyer:1982ah,ref:Becattini,Ambrus:2014uqa,Ambrus:2015lfr,Manning:2015sky} and for interacting fermions. The rotation of interacting fermions was studied in effective field-theoretical approaches~\cite{Chen:2015hfc,Jiang:2016wvv,Ebihara:2016fwa,Chernodub:2016kxh}, in the holographic models~\cite{ref:McInnes} and with the help of numerical methods in the context of Euclidean lattice QCD~\cite{Yamamoto:2013zwa}. In most mentioned cases it assumed that the rotation is rigid so that the angular velocity does not depend on the distance to the axis of rotation.

The basic aim of this paper is to determine how strongly the phase structure of relativistically rotating interacting fermions depends on the choice of the boundary conditions which inevitably appear in the problem of rigid rotation. Indeed, any uniformly rotating system must be bounded in the directions transverse to the axis of rotation in order for the velocity of particles to be smaller than the speed of light~\cite{Davies:1996ks}. Consequently, the fermions must be confined inside a cylinder of sufficiently small radius by appropriate boundary conditions. If no boundary condition were imposed then the fermions would propagate to the causality-violating region where the rotational velocity exceeds the speed of light. The latter inevitably leads to unphysical instabilities and pathological excitations of the rotating system~\cite{Ambrus:2014uqa,ref:Levin,Davies:1996ks,Duffy:2002ss}.

Thus, rapidly rotating systems should necessarily be bounded in the plane transverse to the axis of rotation. As a consequence, they should be subjected to certain finite geometry effects which lead to two effects:
\begin{itemize}
\item[(i)] discretization of the radial component of the particle energy spectrum;
\item[(ii)] appearance of dependence of the energy spectrum on the type of the boundary conditions. 
\end{itemize}
We show in our paper that these effects lead to noticeable quantitative consequences for the phase structure of rotating fermionic systems in thermal equilibrium.

In our previous paper~\cite{Chernodub:2016kxh} we studied properties of interacting rotating matter in a cylinder subjected to the MIT boundary conditions. The main feature of the phase diagram is that the rotation lowers the critical temperature of the chiral symmetry restoration. This finding qualitatively agrees with the results of Ref.~\cite{Jiang:2016wvv} in which the boundaries and, consequently, finite-volume effects of the rotating system were not taken into account. In the present paper we extend our studies to a generalized (chiral) MIT boundary condition and show that the quantitative properties of the phase diagram are noticeably affected by the type of the boundary imposed on the system, while the qualitative features remain the same.

The phase structure of this paper is as follows. In Sect.~\ref{sec:free} we briefly outline main properties of free fermions inside rigidly rotating cylindrical cavity following the original results of Ref.~\cite{Ambrus:2015lfr}. We also generalize the results of Ref.~\cite{Ambrus:2015lfr} to the case of uniformly rotating free fermions confined to a cylinder with so-called chiral MIT boundary conditions. These boundary conditions are characterized by a continuous angle parameter~$\Theta \in [0,2\pi)$. In Sect.~\ref{sec:interacting} we obtain phase diagrams of relativistically rotating fermions described by the Nambu--Jona-Lasinio model for different chiral angles $\Theta$. Following the approach developed in our previous work~\cite{Chernodub:2016kxh} we demonstrate the essential dependence of the phase diagram on the type of the boundary condition. In addition, we find that the critical temperature has a simple quadratic dependence $T_c (\Omega) = T_c^{(0)} - C \,\Omega^2$ on the angular frequency $\Omega$ where both $T_c^{(0)}$ and the positive prefactor $C$ are functions of the chiral angle $\Theta$. The last section is devoted to conclusions.

\section{Rigid rotation of free fermions}
\label{sec:free}

\subsection{Dirac equation in rotating spacetime}

Let us consider a free fermionic particle in a space which rotates with the constant angular velocity $\Omega$ about the fixed $z \equiv x_3$ axis. We assume that all spatial regions of the system rotate uniformly with the same angular velocity (``rigid rotation'') which immediately implies that the size of the system in the plane perpendicular to the rotation axis must be finite. Indeed, any point at the distance $\rho$ from the axis rotates with the velocity $v = \Omega \rho$ which should be smaller than the speed of light $c \equiv 1$ in order to preserve the causality.  Therefore $\Omega \rho \leqslant 1$.

For the sake of convenience we consider the problem in the cylindrical coordinates 
\beqn
x \equiv (x_0, x_1,x_2,x_3) = (t,\rho\sin\varphi,\rho\cos\varphi, z)
\eeqn
corresponding to the reference frame which (co)rotates together with the system at the same angular frequency $\Omega$. The coordinates of the corotating reference frame $x^\mu$ and in the nonrotating laboratory reference frame $x^\mu_\lab$ are related as follows: $t = t_{\lab}$, $\rho = \rho_\lab$, $z = z_\lab$ and 
\beqn
\varphi = [\varphi_{\mathrm{lab}} - \Omega t]_{2\pi}\,,
\label{eq:varphi}
\eeqn
where $[f]_{2\pi}$ corresponds to $f$ modulo $2\pi$. Without loss of generality we assume that the system rotates in the counterclockwise direction, $\Omega \geqslant 0$. The causality requirement restricts the maximal value of the cylinder radius:
\beqn
\Omega R \leqslant 1\,.
\label{ref:bound}
\eeqn

In the corotating frame the spacetime metric is curved:
\beqn
& & ds^2 \equiv g_{\mu\nu} dx^\mu dx^\nu 
\label{eq:metric} \\
& &= \left(1-\rho ^2 \Omega ^2 \right)dt^2- 2\rho^2\Omega dt d\varphi - d\rho ^2- \rho^2 d\varphi^2 - d z^2.
\nonumber
\eeqn
Therefore the rotating fermions of the mass $M$ can be described by the Dirac equation in the curved spacetime:
\beqn
\left[ i \gamma^\mu \left(\partial_\mu + \Gamma^\mu \right)- M \right] \psi = 0\,,
\label{eq:Dirac}
\eeqn
where $\gamma^\mu = \gamma^{\hat \alpha} e^{\;\mu}_{\hat \alpha} $ are the Dirac matrices in the corotating frame and the vierbein $e^{\mu}_{\hat \alpha}$ is the ``square root'' of the curved metric~\eq{eq:metric}: $\eta_{\hat\alpha \hat\beta} =e^{\; \mu}_{\hat\alpha}e^{\; \nu}_{\hat\beta}  g_{\mu \nu}$. Here $\eta_{\hat\alpha \hat\beta} \equiv {\mathrm{diag}} (+1,-1,-1,-1)$ is the flat metric of the laboratory frame (with ``hatted'' indices $\hat\alpha$, $\hat\beta$, etc) while the indices without the hats ($\mu,\nu$, etc) correspond to the corotating frame~\eq{eq:metric}. The corotating and laboratory Dirac matrices satisfy the natural anticommutation relations $\{\gamma^\mu,\gamma^\nu\} = 2 g^{\mu\nu}$ and $\{\gamma^{\hat\mu},\gamma^{\hat\nu}\} = 2 g^{{\hat\mu}{\hat\nu}}$.

In Eq.~\eq{eq:Dirac} the affine connection ,
\beqn
\Gamma_\mu = -\frac{i}{4}\omega_{\mu {\hat \alpha} {\hat \beta}}\sigma^{{\hat \alpha} {\hat \beta}}, 
\label{eq:affine:connection}
\eeqn
is expressed via the spin connection
\beqn
\omega_{\mu \hat{\alpha}\hat{\beta}} = g_{\nu\gamma} \, e_{\hat{\alpha}}^{\;\gamma} \left(\partial_{\mu}e^{\;\nu}_{\hat{\beta}}+ \Gamma^{\nu}_{\,\sigma\mu}e^{\;\sigma}_{\hat{\beta}}\right), 
\label{eq:spin:connection}
\eeqn
the Christoffel symbol 
\beqn
\Gamma _{\mu \nu} ^{\lambda} = \frac{1}{2}g^{\lambda\sigma}\left(g_{\sigma \nu,\mu}+g_{\mu\sigma,\nu}-g_{\mu\nu,\sigma}\right)\,,
\eeqn
and the spin matrix:
\beqn
\sigma^{\hat{\alpha}\hat{\beta}} & = & \frac{i}{2}\left[\gamma^{\hat{\alpha}}, \gamma^{\hat{\beta}}\right].
\eeqn

In the rotating spacetime~\eq{eq:metric} the only nonzero component of the affine connection~\eq{eq:affine:connection} is given by
\begin{align}
\Gamma _{t}= -\frac{i}{2} \Omega \, \sigma^{\hat{x}\hat{y}},
\label{eq:Gamma}
\end{align}
while the nonzero components of the vierbein are as follows:
\beqn
e^{\;t}_{\hat{t}}=e^{\;x}_{\hat{x}}=e^{\;y}_{\hat{y}}=e^{\;y}_{\hat{y}}=1,
\quad 
e^{\;x}_{\hat{t}}= y\Omega, 
\quad
e^{\;y}_{\hat{t}}= -x\Omega. 
\qquad
\eeqn
Further details may be found in Refs.~\cite{Ambrus:2015lfr,Chen:2015hfc,Jiang:2016wvv,Ebihara:2016fwa,Chernodub:2016kxh}.

\subsection{Chiral MIT boundary conditions and solutions}

The fermions should not be able to escape the rotating cylinder in order to avoid a violation of causality. To this end it is convenient to impose on the fermion wavefunctions the following chiral MIT condition at the boundary of the cylinder~$\rho = R$:
\beqn
\bigl[i \gamma^\mu n_\mu(\varphi) - e^{- i \Theta \gamma^5} \bigr] \psi(t,z,\rho,\varphi) {\biggl |}_{\rho = R} = 0\,,
\label{eq:BC}
\eeqn
where $n^\mu(\varphi) = (0, \cos\varphi, \sin\varphi, 0)^T$ 
is a spatial vector normal to the cylinder surface, $\gamma^\mu$ are the Dirac matrices and $\Theta$ is a chiral angle which parameterizes the chiral boundary condition~\eq{eq:BC} via the factor
\beqn
e^{- i \Theta \gamma^5} \equiv \cos \Theta - i \gamma^5 \sin \Theta\,.
\eeqn
The chiral MIT boundary condition~\eq{eq:BC} confines the fermions inside the cylindrical cavity because Eq.~\eq{eq:BC} implies that the normal component of the fermion current,
\beqn
j^\mu = {\bar \psi} \gamma^\mu \psi\,,
\label{eq:j:mu}
\eeqn
vanishes at every point of the surface of the cylinder at any value of the chiral angle $\Theta$:
\beqn
j_{\bs n} \equiv  {\bs j} {\bs n} \equiv - j^\mu n_\mu  = 0 \quad \mbox{at} \ \ \rho = R \,.
\label{eq:j:n}
\eeqn
The usual MIT boundary condition corresponds to $\Theta {=} 0$. This case has been considered in details in Ref.~\cite{Ambrus:2015lfr}.

A general solution of the Dirac equation~\eq{eq:Dirac} with the boundary conditions~\eq{eq:BC} in the rotating reference frame~\eq{eq:varphi} with the spacetime metric~\eq{eq:metric} has the following form:
\beqn
U_j (t,z,\rho,\varphi) = \frac{1}{2\pi} e^{- i \tE t + i k_z z} u_j(\rho,\varphi)\,,
\label{eq:U}
\eeqn
where $u_j$ is an eigenspinor characterized by the cumulative index
\beqn
j = (k_z, m, l, \mathrm{sign}\, E)\,, 
\label{eq:j}
\eeqn
which includes the momentum $k_z \in \R$ along the $z$ axis, the quantized angular momentum $m \in \Z$ with respect to the $z$ axis and the radial excitation number $l = 1,2,3, \dots \in \N$.

The only place where the rotational frequency $\Omega$  enters the solution~\eq{eq:U} is the energy in the corotating frame:
\beqn
\tE_j = E_j - \Omega \Bigl(m + \frac{1}{2}\Bigr) \equiv E_j - \Omega \mu_m\,,
\label{eq:Energy}
\eeqn
where
\beqn
E_j \equiv E_{ml}(k_z,M) = \pm \sqrt{k_z^2 + \frac{q_{ml}^2}{R^2} + M^2}\,,
\label{eq:E:j}
\eeqn
is the energy in the laboratory frame and $\mu_m$ is the eigenvalue of the projection of the total angular momentum onto the $z$ axis:
\beqn
{\hat J}_z \psi = \mu_m \psi\,, \qquad \mu_m =  m + \frac{1}{2}\,.
\label{eq:mu:m}
\eeqn
In the Dirac representation :
\beqn
\gamma ^{\hat{t}}=
\begin{pmatrix}
\bbbone & 0 \\
0 & -\bbbone 
\end{pmatrix}
, \quad
\gamma ^{\hat{i}}=
\begin{pmatrix}
0 &\sigma_i \\
-\sigma_i &0 
\end{pmatrix}
,\quad
\gamma ^5=
\begin{pmatrix}
0 & \bbbone \\
\bbbone & 0 
\end{pmatrix}
, \qquad
\eeqn
and the $z$ component of the total angular momentum operator is as follows:
\beqn
\hat{J}_z  =
- i\partial_\varphi + \frac{1}{2}
    \begin{pmatrix}
\sigma_3 & 0 \\
0 & \sigma_3 
\end{pmatrix}
.
\label{eq:hat:J}
\eeqn

The quantity $q_{ml} \equiv q_{ml}(\Theta,MR)$ in the expression of energy in the laboratory frame~\eq{eq:E:j} is the only quantity which contains information about the chiral angle $\Theta$ and, consequently, about the chiral boundary condition~\eq{eq:BC}. The quantized values of $q_{ml}$, which play a role of the radial contribution to energy, can be determined as follows. 

The spinor $u_j$ in the solution~\eq{eq:U} can be represented in terms of the linear combination~\cite{Ambrus:2015lfr}:
\beqn
u_j(\rho,\varphi) = b^+ u^+_j(\rho,\varphi)  + b^- u^-_j(\rho,\varphi)\,.
\label{eq:linear}
\eeqn
The four-component Dirac eigenspinors of positive and negative helicities ($\lambda = \pm 1/2$, denoted also as $\lambda = \pm$),
\beqn
u^\lambda_j(\rho, \varphi) = 
\frac{1}{\sqrt{2}}
\left(
\begin{array}{c}
\bE_+ \phi^\lambda_j\\
  2\lambda \frac{E}{|E|} \bE_- \phi^\lambda_j
\end{array}
\right),
\label{eq:u:j}
\eeqn
are expressed via the two-component Weyl spinors
\beqn
 \phi^\lambda_j(\rho, \varphi) = \frac{1}{\sqrt{2}}
\left(
\begin{array}{c} 
\bp_{\lambda }e^{im\varphi} J_m(q\rho)\\
  2i\lambda \bp_{-\lambda} e^{i(m+1)\varphi} J_{m+1}(q\rho)
\end{array}
\right),
\label{eq:phi:j}
\eeqn
where
\beqn
 \bp_\pm = \sqrt{1 \pm \frac{k_z}{p}}, \quad
 \bE_\pm = \sqrt{1 \pm \frac{M}{E}}, \quad 
 p= \sqrt{k^2_z + \frac{q^2}{R^2}}.\qquad
\eeqn

Substituting the eigenmode~\eq{eq:U} and \eq{eq:linear} into the boundary condition~\eq{eq:BC} as $\psi \equiv U_j$ and using the explicit form of the eigenspinors~\eq{eq:u:j} we get the following matrix equation for the coefficients $b^\pm$ of the solution~\eq{eq:linear}:
\begin{widetext}
\beqn
i 
\begin{pmatrix}
b^+ \frac{E}{|E|} \bE_-\sigma ^\rho \phi^+ - b^- \frac{E}{|E|} \bE_- \sigma ^\rho \phi^- \\
-b^+ \bE_+\sigma ^\rho \phi^+ -b^- \bE_+\sigma ^\rho \phi^-
\end{pmatrix}
=
\begin{pmatrix}
-\cos \Theta \left( b^+ \bE_+ \phi ^+ + b^-  \bE_+ \phi ^- \right) +i \sin \Theta \left(b^+ \frac{E}{|E|} \bE_- \phi ^+ -b^- \frac{E}{|E|} \bE_- \phi ^- \right) \\
-\cos \Theta \left(b^+ \frac{E}{|E|} \bE_- \phi ^+ -b^- \frac{E}{|E|} \bE_- \phi ^-\right) +i \sin \Theta \left(  b^+ \bE_+ \phi ^+ + b^- \bE_+ \phi ^- \right) 
\end{pmatrix}
\qquad
\label{eq:BC:explicit}
\eeqn
with $\sigma^{\rho}= \sigma_1\cos \varphi + \sigma_2\sin\varphi $. The matrix equation can be reduced to
\beqn
\begin{pmatrix}
\left(\cos \Theta \bE_+ {-} i\sin \Theta \frac{E}{|E|} \bE_-\right)\bp_+J_m {-} \frac{E}{|E|}\bE_-\bp_-J_{m+1} 
& \left(\cos \Theta \bE_+ {+} i\sin \Theta \frac{E}{|E|} \bE_-\right)\bp_-J_m {-} \frac{E}{|E|}\bE_-\bp_+J_{m+1} \\[3mm]
\left(\cos \Theta \bE_+ {-} i\sin \Theta \frac{E}{|E|} \bE_-\right)\bp_-J_{m+1} {+} \frac{E}{|E|}\bE_-\bp_+J_{m} 
& -\left(\cos \Theta \bE_+ {+} i\sin \Theta \frac{E}{|E|} \bE_-\right)\bp_+J_{m+1} {-} \frac{E}{|E|}\bE_-\bp_-J_{m}
\
\end{pmatrix}
\!\!
\begin{pmatrix}
\\[-3mm]
b^+ \\[3mm]
b^-\\[1mm]
\end{pmatrix}
{=} 0.
\qquad
\label{eq:BC:explicit:2}
\eeqn
\end{widetext}

We find that the boundary condition~\eq{eq:BC:explicit:2} has a nontrivial solution in terms of the coefficients $b^{\pm}$ provided the quantity $q$ satisfies the following relation:
\beqn
\cos \Theta\left(\jj_{m}^2(q) + \jj_m(q) \, \frac{2 M R}{q} \cos\Theta -1 \right) = 0\,,
\eeqn
or
\beqn
\jj_{m}^2(q) + \jj_m(q) \, \frac{2 M R}{q} \cos\Theta -1  = 0\,,
\label{eq:jj}
\label{eq:J}
\eeqn
where
\beqn
\jj_m(x) = \frac{J_m(x)}{J_{m+1}(x)}\,,
\label{eq:jjm}
\eeqn
is the ratio of the Bessel functions $J_m(x)$. Thus, the dimensionless quantity $q_{ml}$ in Eq.~\eq{eq:E:j} is the $l^{\mathrm{th}}$ positive root $(l=1,2, \dots )$ of Eq.~\eq{eq:J}. 

\begin{figure}
\includegraphics[scale=0.5,clip=true]{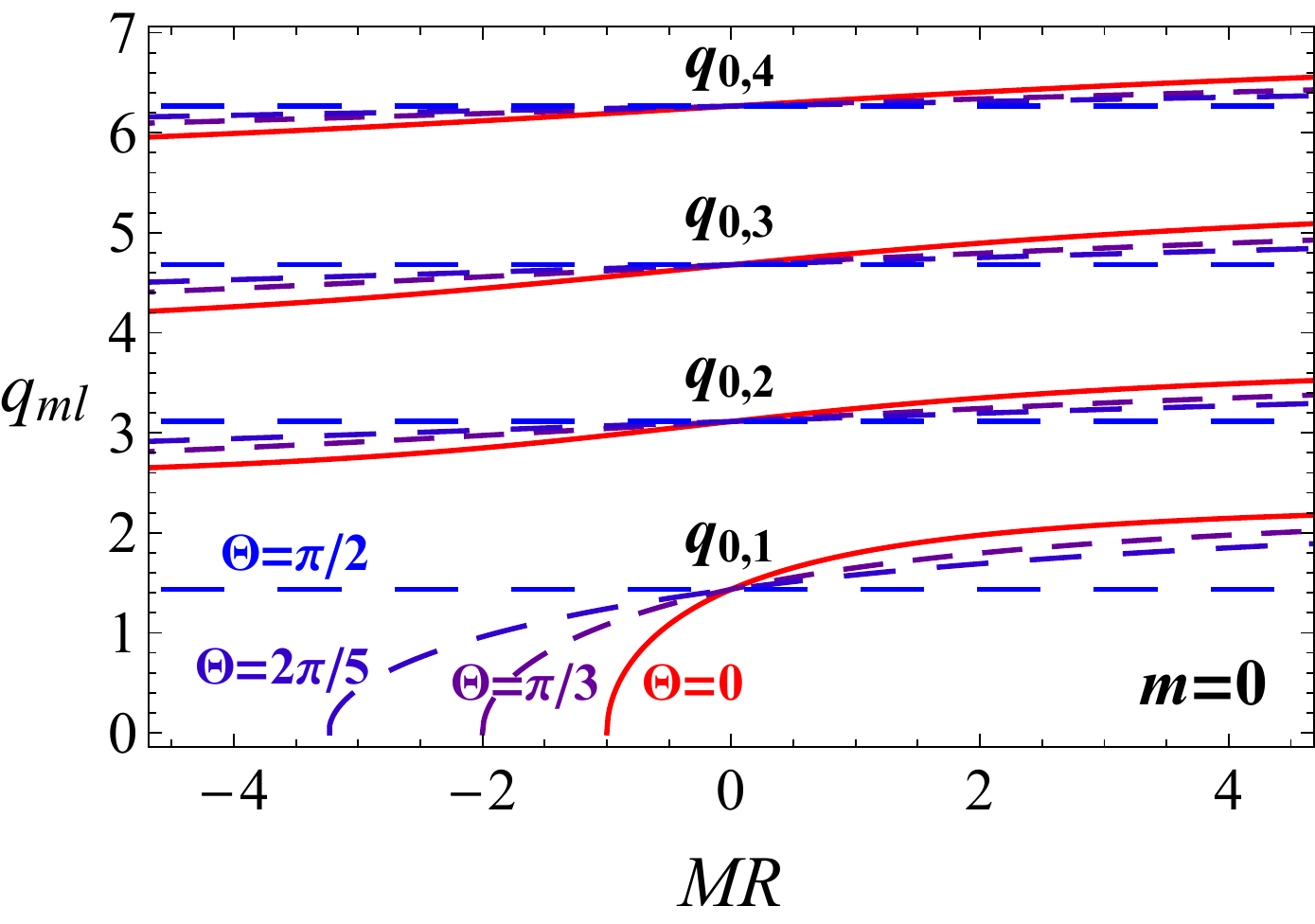}
\vskip -3mm
\caption{Four lowest branches ($l=1,\dots,4$) of the $m=0$ solution $q_{ml}$ of Eq.~\eq{eq:J} vs the normalized mass~$M R$ for a range of chiral boundary parameters $\Theta = 0, \pi/3, 2\pi/5, \pi/2$. }
\label{fig:q:mls}
\end{figure}

The angle $\Theta$, which determines the boundary condition~\eq{eq:BC}, enters the eigenenergies~\eq{eq:Energy} and \eq{eq:E:j} only via the parameter $q$ which is, in turn, determined by Eq.~\eq{eq:jj}. In the special cases of the boundary condition, $\Theta = 0, \pi$, Eq.~\eq{eq:jj} coincides with the known relations~\cite{Ambrus:2015lfr}. According to Fig.~\ref{fig:q:mls} the boundary parameter $\Theta$ affects the spectrum of $q$'s quite significantly. Notice that at the special value $\Theta = \pi/2$ the values of $q_{ml}$ are independent of the mass $M$. 

The solutions $q_{ml}$ with the fixed radial $l=1$ and angular $m\in \Z$ quantum numbers become zero 
\beqn
q_{m1}{\biggl|}_{f_m=0} = 0,
\label{eq:q:zero}
\eeqn
at the quantized values of the mass $M$ satisfying the following condition:
\beqn
0 = f_m \equiv 
\left\{
\begin{array}{lll}
M R\cos \Theta + 1 +m , & \quad & m \geqslant 0  \\
M R\cos \Theta - m , & \quad & m< 0
\end{array}
\right.
\,.
\label{eq:cond:q:zero}
\eeqn
Equations~\eq{eq:q:zero} and \eq{eq:cond:q:zero} can found by expanding the Bessel function around the origin $x= 0$. Since condition~\eq{eq:cond:q:zero} does not hold at $\Theta=\pi /2$, the radial eigenvalue $q_{m1}$ does not touch the $q_{m1}=0$ axis in accordance with Fig.~\ref{fig:q:mls}. Notice that the positive and negative $m$ in condition~\eq{eq:q:zero} are related by the symmetry with respect to the flips of the orbital number $\mu_{m} \equiv - \mu_{-m-1}$ which leaves the eigenvalue $q_{ml}$ unchanged:
\beqn
q_{m,l} = q_{-m-1,l}\,.
\label{eq:qml:eq}
\eeqn

\begin{figure}[!thb]
\begin{center}
\includegraphics[scale=0.4,clip=true]{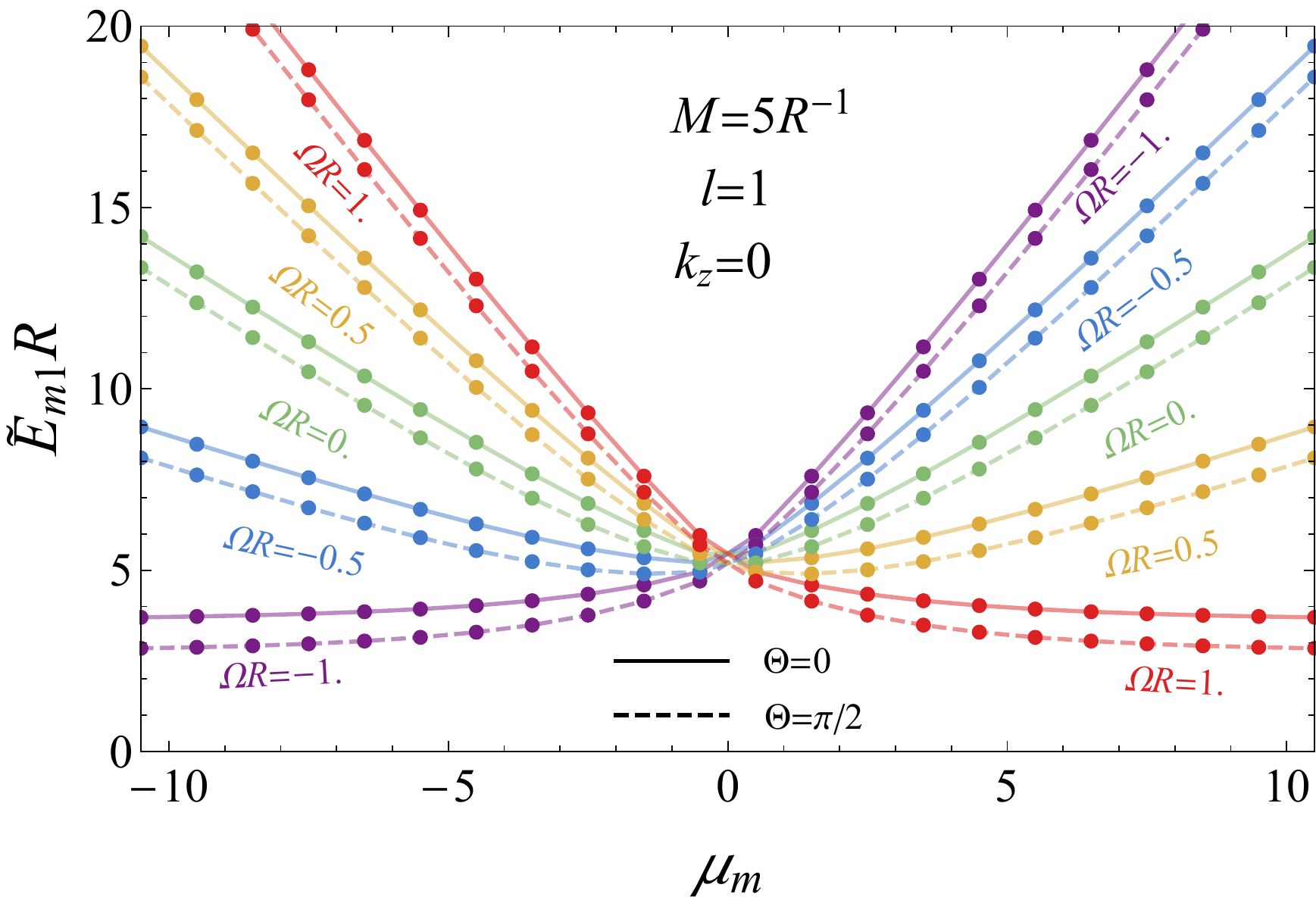}\\[3mm]
\includegraphics[scale=0.4,clip=true]{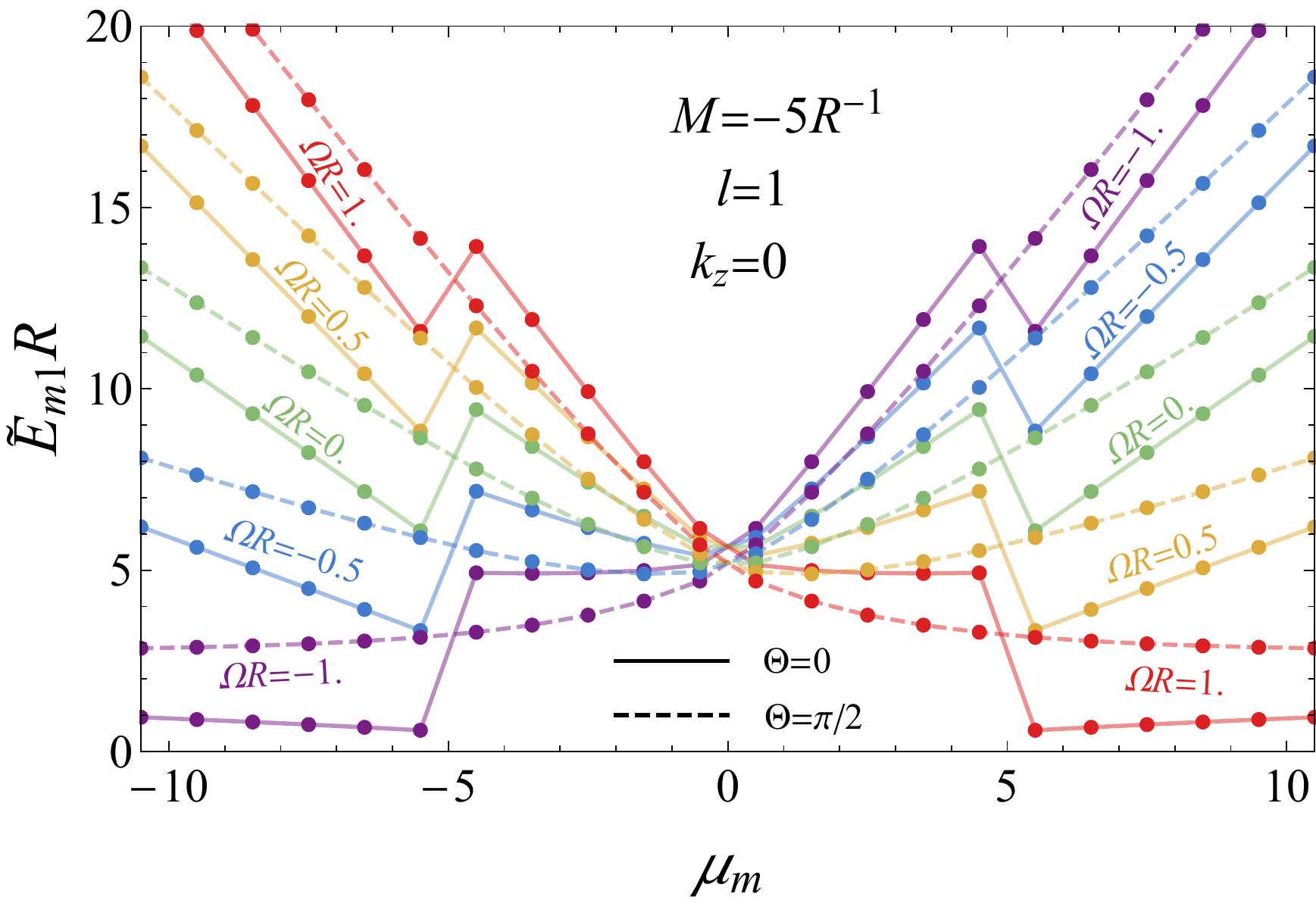}
\end{center}
\vskip -5mm
\caption{Lowest energy eigenmodes (with $l = 1$ and $k_z = 0$) in the corotating frame~\eq{eq:Energy} vs. the total angular momentum $\mu_m$, Eq.~\eq{eq:mu:m}, for the positive mass $MR = 5$ (the upper plot) and the negative mass $MR = -5$ (the lower plot) for various values of the rotation frequency $\Omega$ and two values of the boundary parameter $\Theta = 0 $ (the solid lines) and $\Theta = \pi/2$ (the dashed lines).}
\label{fig:emum}
\end{figure}

The energy spectrum of the model has a noticeable dependence on $\Theta$. In Fig.~\ref{fig:emum} we plot the lowest eigenvalue of the energy in the corotating frame as a function of the $z$-component of the total angular momentum $\mu_m$ for two values of the boundary parameters: $\Theta = 0$ and $\Theta = \pi/2$. The former value corresponds to the MIT boundary condition while at the latter value $q_{ml}$ becomes independent of the mass. We show that the boundary angle $\Theta$ affects the spectrum both at positive and negative values of the mass $M$. Thus we may expect that the phase diagram of the interacting fermions will exhibit a certain dependence on the boundary condition.

We would like to make a short remark concerning boundary conditions other than the chiral extension of the MIT conditions~\eq{eq:BC}. It seems plausible that for rotating states a physical boundary condition should be implemented in an essentially local way at every point of the cylinder [as it is done, for example, in Eq.~\eq{eq:j:n}]. A local condition should be contrasted to a general integral boundary condition implemented nonlocally at the whole surface $S \equiv \partial V$ which bounds the volume $V$ of the rotating cylinder. For instance, an integral condition of the form~\cite{Ambrus:2015lfr,Ebihara:2016fwa}
\beqn
\oint_{S} j^\mu(x) d \Sigma_\mu = 0\,,
\label{eq:boundary:integral}
\eeqn
ensures the conservation of the global fermion number inside the cylinder, and, at the same time it does not prevent individual fermions from escaping the cylinder and coming back in such a way that the global condition~\eq{eq:boundary:integral} is still preserved. In other words, certain escaped fermions will travel through the region beyond the ``light surface'' LS with the radius $R_{\mathrm{LS}} = \Omega^{-1}$. It is these fermions which cause -- after travelling through the causality-violation region at $R > R_{\mathrm{LS}}$ -- known unphysical instabilities and excitations of the rigidly rotating system~\cite{ref:Levin,Davies:1996ks,Duffy:2002ss}.

Concluding this section, we notice that in the cylindrical volume the integration over the three momentum~$\bs k$ is modified with respect to the one in an unbounded space~\cite{Ambrus:2015lfr}:
\beqn
\int \frac{d^3 k}{(2 \pi)^3} \to \sum_j \equiv \frac{1}{\pi R^2}\sum_{l=1}^\infty \sum_{m= -\infty}^\infty \int \frac{d k_z}{2 \pi} \,.
\label{eq:phase:space:k}
\eeqn

\section{Interacting fermions in rotation}
\label{sec:interacting}

\subsection{Free energy}

In our paper we consider a simplest description of interacting fermions which is given by the Nambu--Jona-Lasinio (NJL) model~\cite{ref:NJL}. In the corotating reference frame the action of the NJL  for a single massless (chiral) fermion species is given by the following formula:
\beqn
S_{\mathrm{NJL}} & = & \int_{\mathrm{V}} d^4x \sqrt{- \mathrm{det}\left(g_{\mu \nu}\right)} \, \cL_{\mathrm{NJL}}\left(\bar{\psi},\psi\right), \label{eq:L:NJL} \\
\cL_{\mathrm{NJL}} & = & \bar{\psi} i\gamma _\mu\left(\partial^\mu + \Gamma ^\mu \right)\psi +\frac{G}{2}\left[\left(\bar{\psi}\psi\right)^2+ \left(\bar{\psi}i\gamma_5 \psi \right)^2 \right]\,,
\nonumber
\eeqn
where $g_{\mu\nu}$ corresponds to the metric~\eq{eq:metric} and $\Gamma^\mu$ is the connection~\eq{eq:Gamma}. Then we perform a Hubbard-Stra\-to\-no\-vich transformation by introducing an auxiliary sca\-lar sigma field ($\sigma \sim G \avr{{\bar q} q}$) and a pseudoscalar ``pion'' field ($\pi \sim G \avr{{\bar q} i \gamma^5 q}$). In the ground state of our $CP$-invariant system $\pi \equiv 0$. The ground state is determined by the condensate of the sigma field which is assumed to be uniform.\footnote{In a striking difference with the background magnetic field, the relativistic rotation may lead to inhomogeneities of the condensate which make the analytical treatment of the problem very difficult. At the moment it is customary to assume that $\sigma$ is a coordinate-independent quantity~\cite{Chen:2015hfc,Jiang:2016wvv,Chernodub:2016kxh}.
}

Skipping all steps of the derivation that can be found in detail in Ref.~\cite{Chernodub:2016kxh}, we arrive to the following expression for the density of the (Helmholtz) free energy in the corotating frame:
\beqn
{\widetilde F}(\sigma)  & = & \frac{\sigma^2}{2G} + V_{\mathrm{vac}}(\sigma) + V_{\mathrm{rot}}(\sigma;T,\Omega)\,.
\label{eq:F:free:energy}
\eeqn
The vacuum part of the potential,
\beqn
V_{\mathrm{vac}}(\sigma) & = & - \frac{1}{\pi R^2} \sum_{m \in \Z} \sum\limits_{l=1}^\infty \int \frac{d k_z}{2 \pi}\nonumber \\
& & 
\cdot f_\Lambda\left(\sqrt{k_z^2 + \frac{q_{ml}^2(\sigma)}{R^2}}\right) E_{ml}(k_z,\sigma),
\label{eq:V:vac:reg}
\label{eq:V:vacuum}
\eeqn
is divergent in the ultraviolet limit and therefore it has to be regularized. In our paper we take the following phenomenological cutoff function~\cite{Gorbar:2011ya,Chen:2015hfc}:
\beqn
f^{{\mathrm{exp}}}_\Lambda(\varepsilon) = \frac{\sinh(\Lambda/\delta \Lambda)}{\cosh(\varepsilon/\delta \Lambda) + \cosh (\Lambda/\delta \Lambda)},
\label{eq:f:Lambda:2}
\eeqn
with the value of the cutoff $\delta \Lambda = 0.05 \, \Lambda$. In Equation~\eq{eq:V:vac:reg} the energy $E_{ml}$ is given in Eq.~\eq{eq:E:j}:
\beqn
E_j \equiv E_{ml}(k_z,M) = \pm \sqrt{k_z^2 + \frac{q_{ml}^2}{R^2} + \sigma^2}\,,
\label{eq:E:j:sigma}
\eeqn
while the value of $q_{ml}$ is determined by Eq.~\eq{eq:jj},
\beqn
\jj_{m}^2(q) + \jj_m(q) \, \frac{2 R}{q} \sigma \cos\Theta -1  = 0\,,
\label{eq:q:sigma}
\eeqn
where we identified $M$ with $\sigma$. The functions $\jj_{m}$ are defined by Eq.~\eq{eq:jjm}. 

The rotational contribution to the free energy~\eq{eq:F:free:energy} is given by the following formula:
\beqn
& & V_{\mathrm{rot}}(\sigma;T,\Omega) = - \frac{T}{\pi R^2} \sum_{m \in \Z}  \sum\limits_{l=1}^\infty \int \frac{d k_z}{2 \pi} \\
& & \qquad \quad 
\cdot \biggl[\ln \left( 1 + e^{- \frac{E_{ml}(k_z,\sigma) - \Omega \mu_m}{T}} \right) \nonumber \\
& & \qquad \quad \ \ + \ln \left( 1 + e^{-\frac{E_{ml}(k_z,\sigma) +\Omega \mu_m}{T}} \right),
\biggr], \qquad
\nonumber
\eeqn
where the energy~$E_{ml}$ and the total angular momentum~$\mu_m$ are given in Eqs.~\eq{eq:Energy} and \eq{eq:mu:m}, respectively, and
the values of $q_{ml}$ are again determined by Eq.~\eq{eq:q:sigma}.

The thermodynamic properties of the system of interacting fermions are determined by the free energy~\eq{eq:F:free:energy}. In the mean-field approach the ground state is given by a global minimum of the free energy~\eq{eq:F:free:energy}. If the corresponding  solution is nontrivial, then the spontaneous mass gap generation takes place. In this phase the chiral symmetry is no more respected as the field $\sigma$ plays a role of the dynamical mass for the fermionic field $\psi$.

\subsection{Finite-geometry effects}

We study the properties of interacting fermions in the cylindrical geometry. This system has an infinite volume because of the infinite extension of the space in the $z$ direction.Therefore it cannot be, formally speaking, called a finite-volume system. On the other hand, the effects of the finite radius of the cylinder are quite essential (especially in the rigidly rotating case) and similar to the effects of the finite volume. We therefore prefer to use the notion of a finite-geometry system.

In the case of the MIT boundary conditions $\Theta = 0$, the finite geometry with the MIT boundary conditions affects strongly the phase structure of the model. In Ref.~\cite{Chernodub:2016kxh} we have found that there are three distinct regions in the phase diagram characterized by nature and strength of chiral symmetry breaking. There are explicitly broken (gapped), partially restored (nearly gapless) and spontaneously broken (gapped) phases at, respectively, small, moderate and large radius of the cylinder. Due to the presence of the boundary the chiral condensate experiences specific steplike discontinuities as the function of the coupling constant $G$, temperature $T$ and angular frequency $\Omega$. We notice that these steplike discontinuities have the same nature as the Shubnikov--de Haas oscillations~\cite{ref:SdH} with the exception that they occur in the absence of external magnetic field and Fermi surface. At finite temperature the rotation leads to restoration of spontaneously broken chiral symmetry so that as the temperature increases the critical angular frequency decreases~\cite{Jiang:2016wvv,Chernodub:2016kxh}. At zero temperature the vacuum is insensitive to rotation (``cold vacuum cannot rotate")~\cite{Ebihara:2016fwa,Chernodub:2016kxh}.  Moreover, one can show that at fixed temperature the chirally restored phase possesses a higher moment of inertia compared to the chirally broken phase~\cite{Chernodub:2016kxh}.

Below we consider the case at general value of the angle $\Theta$ of the chiral MIT boundary conditions.
According to the structure of the energy levels~\eq{eq:E:j:sigma} and \eq{eq:q:sigma}, the free energy is invariant under the following flips of the chiral boundary angle $\Theta \in [0, 2\pi)$ with simultaneous (in needed) change of the sign of the condensate:
\beqn
\begin{array}{rcl}
\Theta & \to \pi - \Theta\,, \qquad & \sigma \to - \sigma\,,\\[1mm]
\Theta & \to 2\pi - \Theta\,, \qquad & \sigma \to \sigma\,,
\end{array}
\label{eq:flip:1}
\eeqn
so that the following relations of the free energy in the corotating frame hold:
\beqn
\begin{array}{rcl}
{\widetilde F}(\sigma,\Theta) & = & {\widetilde F}(-\sigma, \pi - \Theta)\,,\\[1mm]
{\widetilde F}(\sigma,\Theta) & = & {\widetilde F}(\sigma, 2\pi - \Theta)\,.
\end{array}
\label{eq:flip:2}
\eeqn
Therefore we show only the values of the chiral boundary angle in the interval $\Theta \in [0,\pi/2]$ while other values of $\Theta$ can be restored from the symmetries~\eq{eq:flip:1} or~\eq{eq:flip:2}.

First of all, in order to disentangle the effects of finite geometry and rotation we consider the ground state of the model in the absence of rotation, $\Omega = 0$. Following our previous study~\cite{Chernodub:2016kxh} we work at the fixed value of the NJL coupling $G = 42 / \Lambda^2$ and the radius of the cylinder $R = 20/\Lambda$. Moreover, the effects of rotation may only appear at nonzero temperature as it is seen from the free energy~\eq{eq:F:free:energy}. Therefore we take $T=0.33 \Lambda$ (the choice of this value will be clear below).

\begin{figure}[!thb]
\begin{center}
\includegraphics[scale=0.4,clip=true]{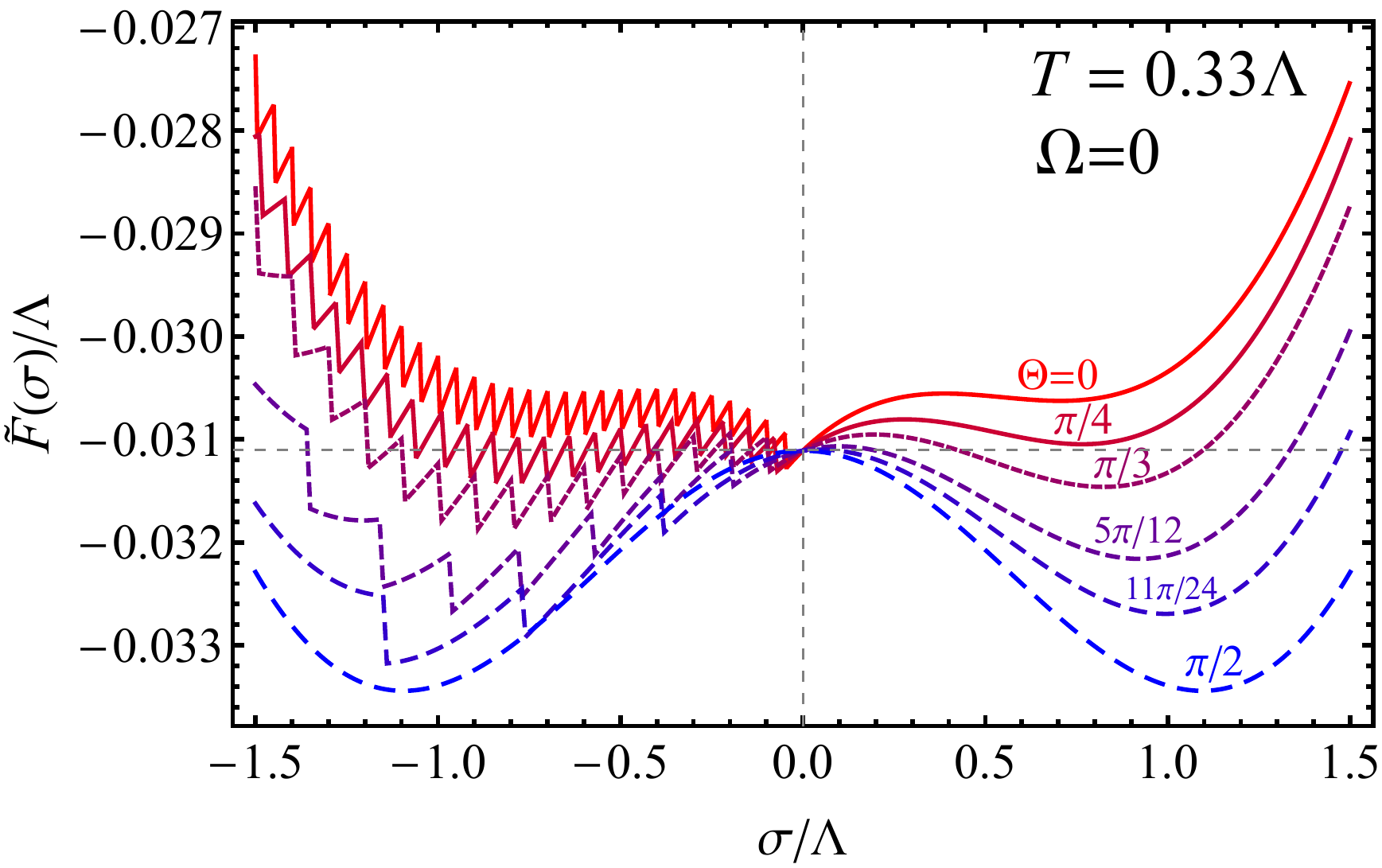}
\end{center}
\caption{Free energy \eq{eq:F:free:energy} as function of the field $\sigma$ in static system with $\Omega = 0$ at temperature $T=0.33 \Lambda$, radius of the cylinder $R = 20/\Lambda$, and various values of the boundary angle $\Theta$.}
\label{fig:potentials:Omega0}
\end{figure}

In Fig.~\ref{fig:potentials:Omega0} we plot the free energy in the corotating frame vs the field $\sigma$ at various values of the boundary angle~$\Theta$ of the nonrotating fermionic matter $\Omega=0$. 

The ground state condensate corresponds to the minimum of the free energy. The plot in Fig.~\ref{fig:potentials:Omega0} demonstrates the following features of the system, most of which are, in fact, generic for all values of the radius of the cylinder $R$, for all temperatures $T$ and couplings $G$:
\begin{enumerate}

\item At low values of the boundary angle $\Theta$ the system resides in a phase with a dynamically unbroken phase in which a weak explicit violation of the chiral symmetry occurs. This slightly broken phase is characterized by a small value of the condensate $\sigma = - 1/(R \cos \Theta)$ which corresponds to the eigenvalue of $q_{01}$ determined by Eqs.~\eq{eq:q:zero} and \eq{eq:cond:q:zero}. The explicit breaking is caused by a proximity effect of the boundary conditions~\eq{eq:BC} which are not invariant under the chiral symmetry~\cite{Chernodub:2016kxh}. 

\item For all values of $\Theta$ (except for the special case $\Theta = \pi/2$) the free energy is not invariant under the flips of the sign of the condensate $\sigma \to - \sigma$. This happens because the variable $q_{ml}$ in the energy of the system~\eq{eq:E:j:sigma}, according to Eq.~\eq{eq:q:sigma}, is not invariant under these flips unless $\Theta = \pi/2$.

\item According to Eq.~\eq{eq:flip:1} at $ \Theta < \pi/2$ and $\Theta > 3\pi/2$ the ground state is given by a negative value of $\sigma$ while at $3\pi/2 > \Theta > \pi/2$ the ground state value $\sigma$ is positive. At $\Theta = \pi/2$ and $ \Theta = 3\pi/2$ the potential is symmetric and the ground state is double degenerate. 

\item  At $\Theta = \Theta_c \approx 5 \pi/24$ a first-order chiral transition occurs: the ground state condensate $\sigma$ suddenly changes from a small negative quantity to a larger negative quantity implying that at $\Theta>\Theta_c$ the mass gap is generated dynamically. Notice that the quoted value of $\Theta_c$ is specific for our chosen set of parameters $T=0.33 \Lambda$, the cylinder radius $R = 20/\Lambda$ and coupling constant $G = 42/\Lambda^2$.

\item As the value of $\Theta$ increases towards $\pi/2$ the minimum increases irregularly its absolute value.

\item The irregularities for the negative values of $\sigma$ appear at $\sigma R \cos \Theta = -1 - m$ for $m= 0, 1, \dots.$ The origin of the irregularities is the zero solutions of $q_{m1}$ given in Eq.(\ref{eq:q:zero}). As $\Theta$ increases towards $\pi/2$, the $m$-th point of the irregularities becomes the larger absolute value of $\sigma,$ and finally at $\Theta=\pi/2$ the irregularities disappear.
\end{enumerate}

\begin{figure}[!thb]
\begin{center}
\includegraphics[scale=0.5,clip=true]{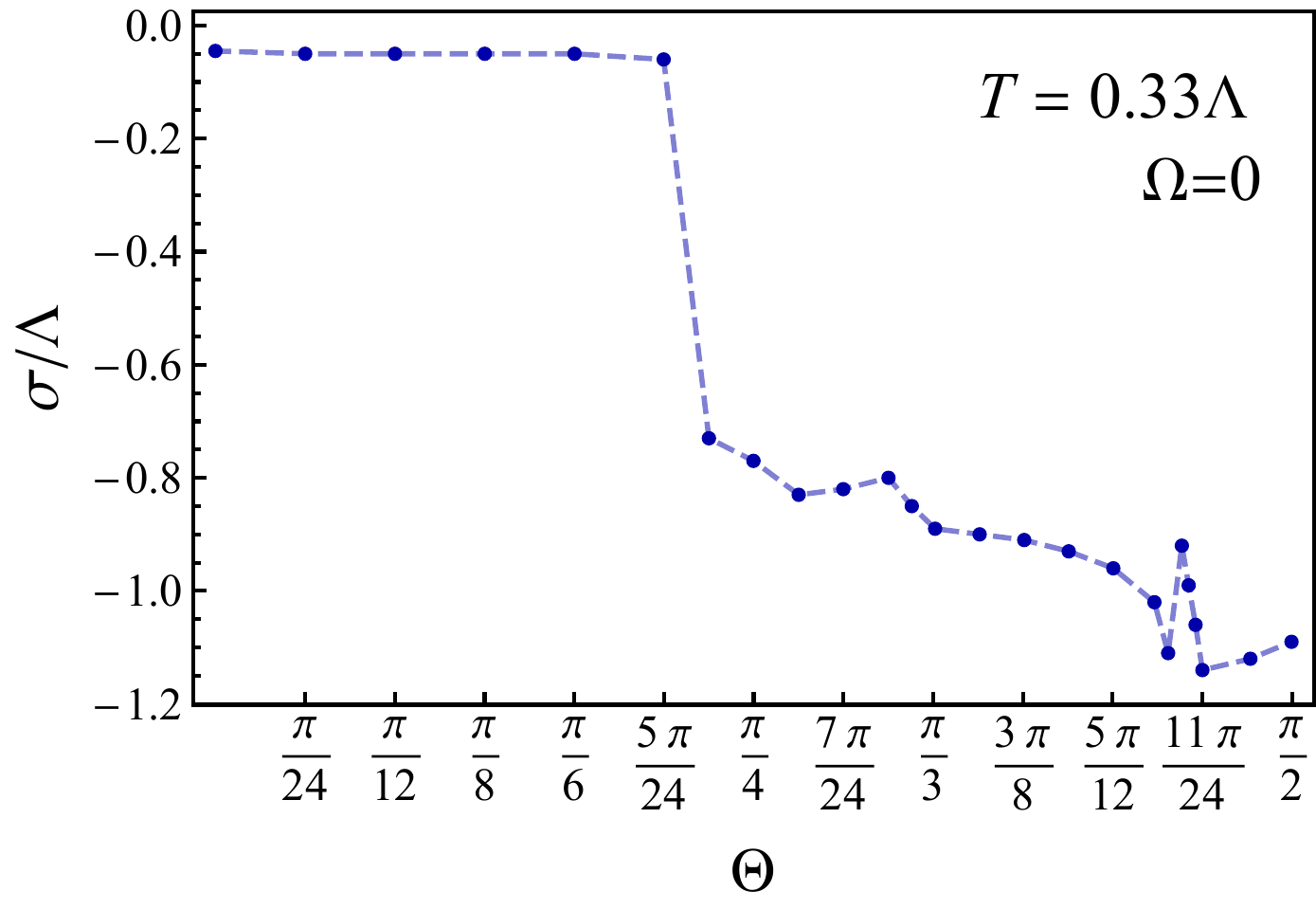}
\end{center}
\caption{The condensate $\sigma$ in the ground state in nonrotating cylinder of the radius $R = 20/\Lambda$ at $T=0.33 \Lambda$ and coupling $G = 42/\Lambda^2$ as a function of the boundary angle $\Theta$. The line is drawn to guide the eye.}
\label{fig:condensate:Omega0}
\end{figure}

In Fig.~\ref{fig:condensate:Omega0} we show the mean-field condensate $\sigma$ as a function of the boundary angle $\Theta$. The model has a clear transition from the chirally restored phase to the dynamically broken chiral phase at $\Theta_c \approx 5 \pi/24$, while in the chirally broken phase the condensate has slightly irregular behavior. According to the symmetry pattern~\eq{eq:flip:1}, as the boundary angle $\Theta$ crosses the value $\Theta = \pi/2$ the chiral condensate should flip the sign.

We conclude that the chiral angle $\Theta$ of the boundary condition affects strongly the phase diagram of interacting fermions in the finite geometry. Below we extend our studies to the case of the rotating fermionic matter.

\subsection{Phase diagrams at different chiral boundaries}

\begin{figure}[!thb]
\begin{center}
\includegraphics[scale=0.427,clip=true]{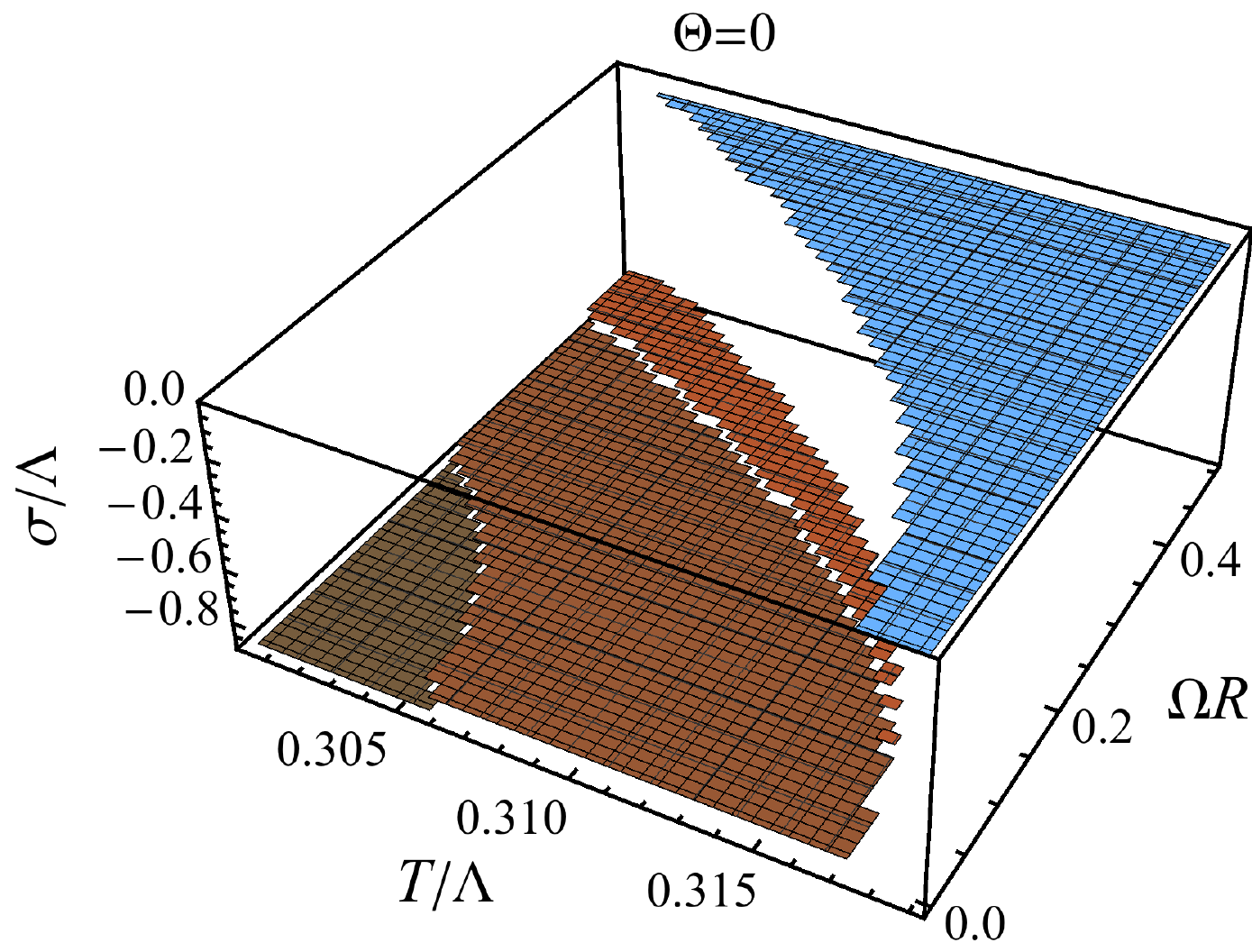} \\[2mm]
\includegraphics[scale=0.427,clip=true]{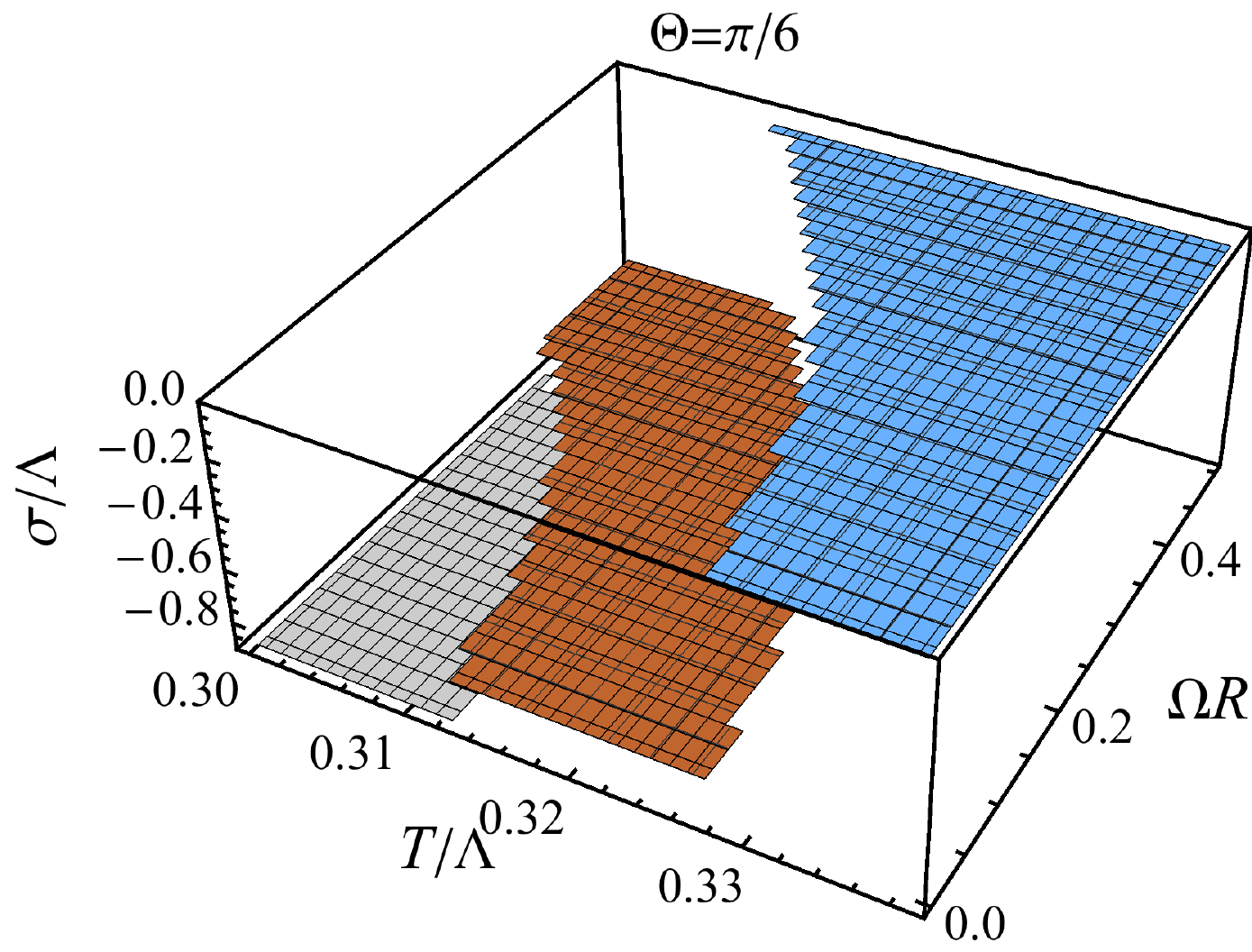} \\[2mm]
\includegraphics[scale=0.427,clip=true]{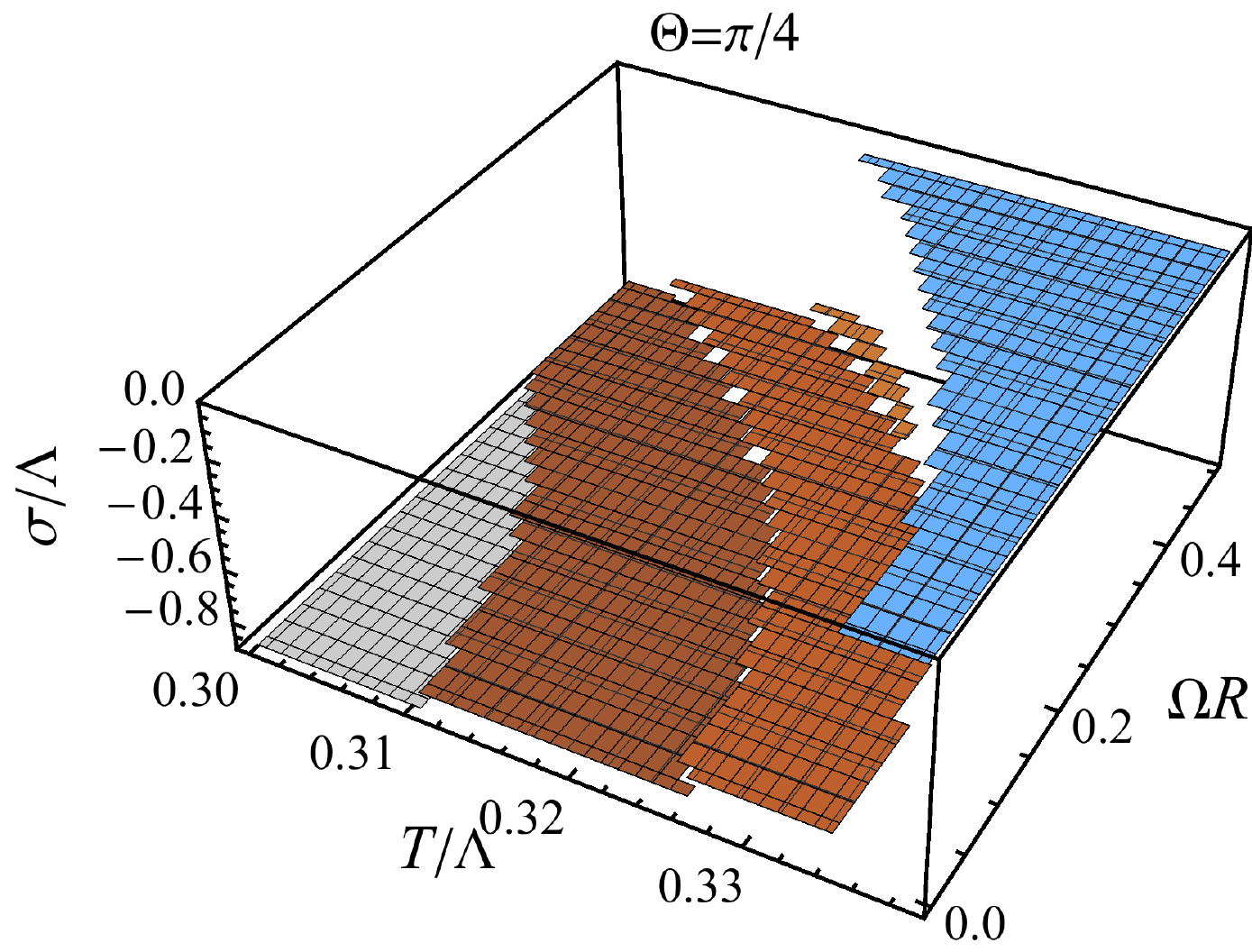} \\[2mm]
\includegraphics[scale=0.427,clip=true]{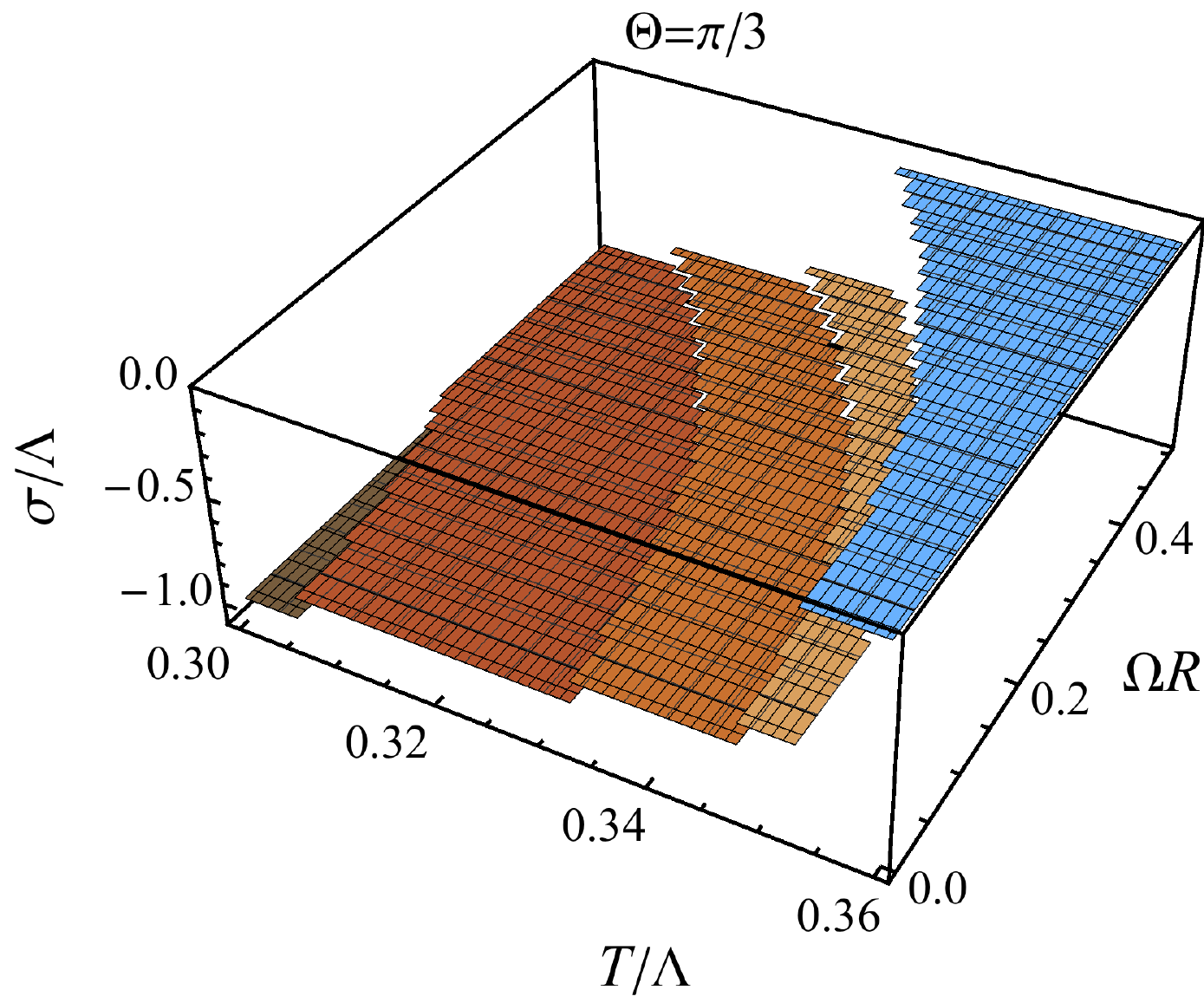} 
\end{center}
\caption{Chiral condensate $\sigma$ in the ``temperature ($T$) -- rota\-ti\-onal frequency ($\Omega$)'' plane for different boundary angles $\Theta$ of the chiral MIT boundary conditions $\Theta = 0, \pi/6, \pi/4, \pi/3$ (from the top to the bottom) at the cylinder radius $R = 20/\Lambda$ and the coupling $G=42/\Lambda^2$.}
\label{fig:sigma:3d}
\end{figure}

In Fig.~\ref{fig:sigma:3d} we plot the chiral condensate in the temperature-angular frequency $(T,\Omega)$ plane for various angles of the boundary angle $\Theta$. We notice several features of the condensate:
\begin{enumerate}

\item[A)] At each chiral boundary angle $\Theta$ the condensate evolves in a series of discontinuous steps between several plateaux. Each plateau has a fixed value of the condensate. This behavior is expected in view of our results of the phase diagram obtained at the standard MIT boundary conditions at $\Omega = 0$ in Ref.~\cite{Chernodub:2016kxh}. The discontinuities appear due to sharp minima of the free energy, which in turn depend on the discontinuous behavior of the radial momenta $q$ as the function of the condensate $\sigma$ (the mass $M$ of the fermion), as illustrated in Fig.~\ref{fig:q:mls} and given in Eq.(\ref{eq:q:zero}). 

\item[B)] The gap between each plateau is both temperature and frequency independent. In other words, the plateaux are flat. 

\item[C)] The plateau with the restored chiral symmetry, shown by the blue flat surface in all plots of Fig.~\ref{fig:sigma:3d}, appears at highest temperatures for all studied values of the boundary chiral angle $\Theta$. 

\item[D)] The critical temperature becomes smaller as the rotational frequency $\Omega$ increases for all $\Theta$. This feature was observed in our previous study at $\Theta = 0$ in Ref.~\cite{Chernodub:2016kxh} and it is also in agreement with Ref.~\cite{Jiang:2016wvv}.

\end{enumerate}

We expect that beyond the mean-field approximation the fluctuations of the $\sigma$ and pion fields $\pi$ should smoothen the discontinuities while the steplike features will surely remain. These steps are (irregular) rotational analogues of the Shubnikov--de Haas oscillations in magnetic field~\cite{Chernodub:2016kxh}. 

\begin{figure}[!thb]
\begin{center}
\includegraphics[scale=0.51,clip=true]{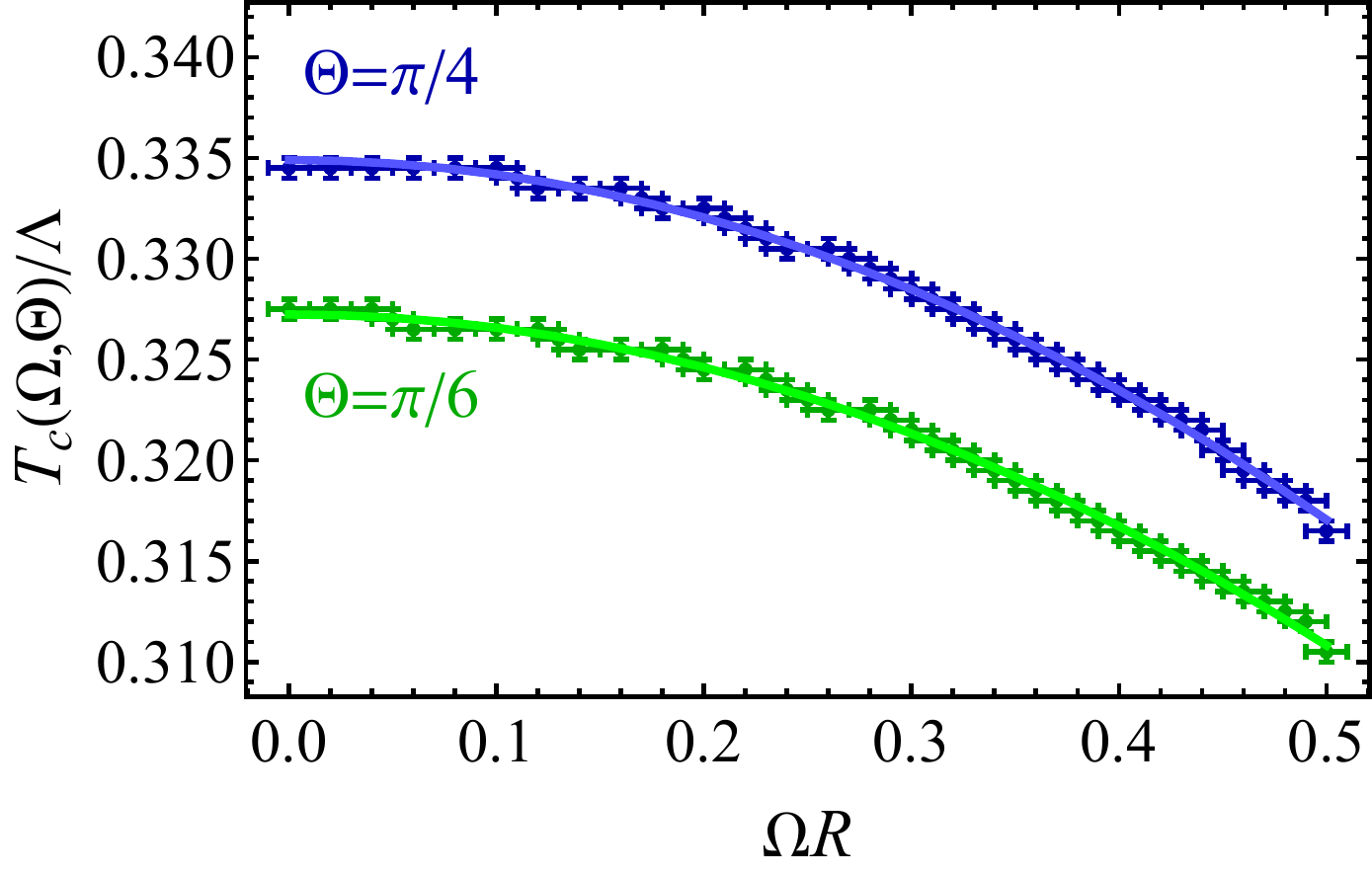}
\end{center}
\caption{Fits of the critical chiral temperature $T_c$ as the function of the angular frequency $\Omega$ for two values of the chiral boun\-dary angle, 
$\Theta=\pi/4$ and $\Theta=\pi/6$. The rotating fermionic matter is placed in the cylinder of the radius $R = 20/\Lambda$ at the coupling $G=42/\Lambda^2$.}
\label{fig:phase:fits}
\end{figure}

We found that for all chiral boundary angles $\Theta$ the critical temperature $T_c = T_c(\Omega,\Theta)$ may be very well described by the quadratic function of the angular velocity~$\Omega$:
\beqn
T_c(\Omega,\Theta) = T^{(0)}_c(\Theta) - C(\Theta) \Omega^2\,.
\label{eq:Tc:fit}
\eeqn
In Fig.~\ref{fig:phase:fits} we show examples of the fits of the critical temperature by the fitting function~\eq{eq:Tc:fit} in which the critical temperature for nonrotating matter $T^{(0)}_c(\Theta)$ and the slope parameter $C(\Theta)$ serve as the fitting parameters at each fixed value of $\Theta$. The fits match the numerical data very well as one can see from the figure.

\begin{figure}[!thb]
\begin{center}
\includegraphics[scale=0.5,clip=true]{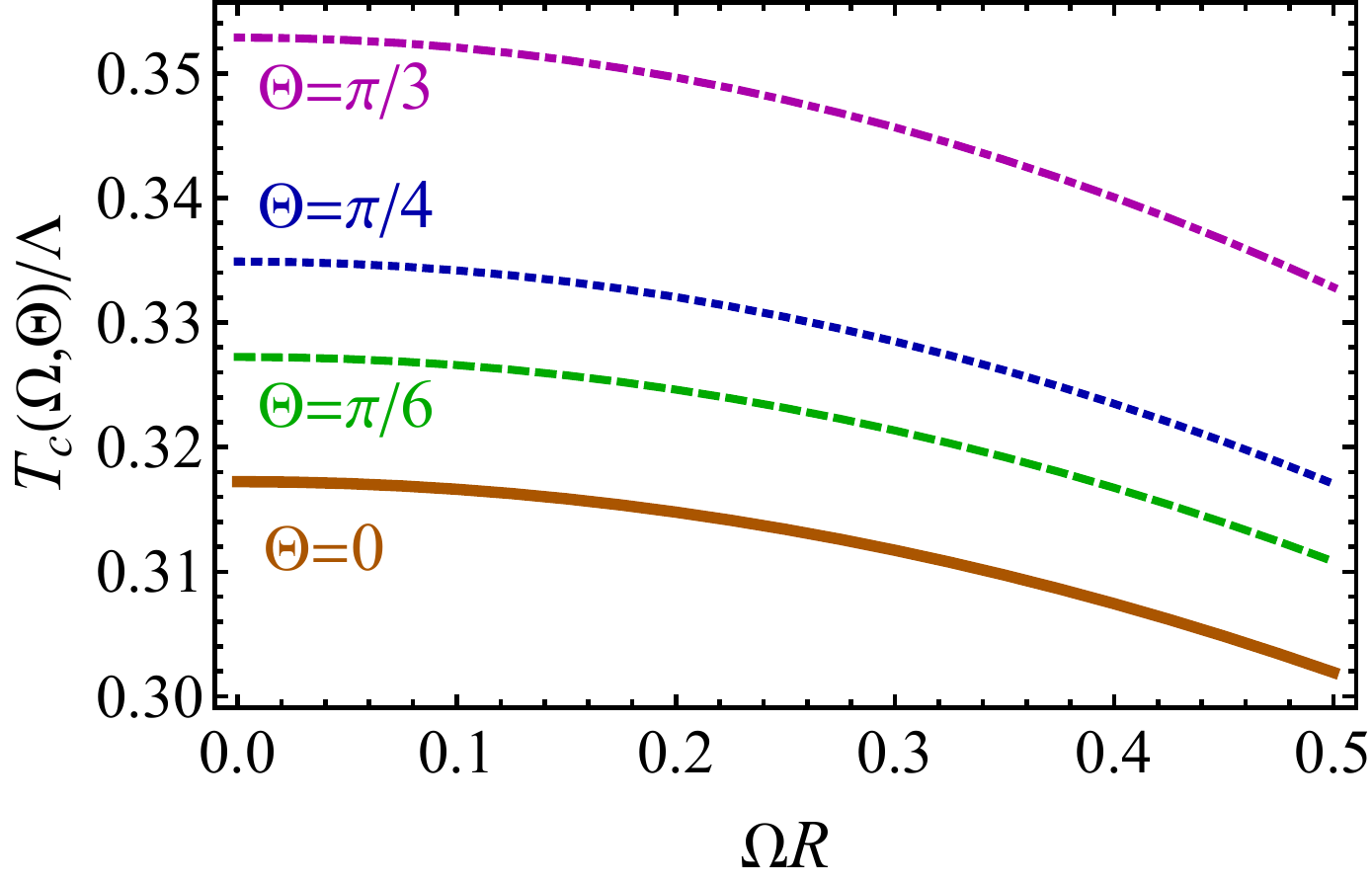}
\end{center}
\caption{Phase diagrams of the rotating fermionic matter in $(T,\Omega)$ plane at different angles $\Theta$ of the chiral MIT boundary condition for the cylinder of the radius $R = 20/\Lambda$ and the coupling $G=42/\Lambda^2$. The dynamically broken chiral symmetry is realized at lower part of the diagram, $T < T_c(\Omega,\Theta)$.}
\label{fig:phase}
\end{figure}

\begin{figure}[!thb]
\begin{center}
\includegraphics[scale=0.49,clip=true]{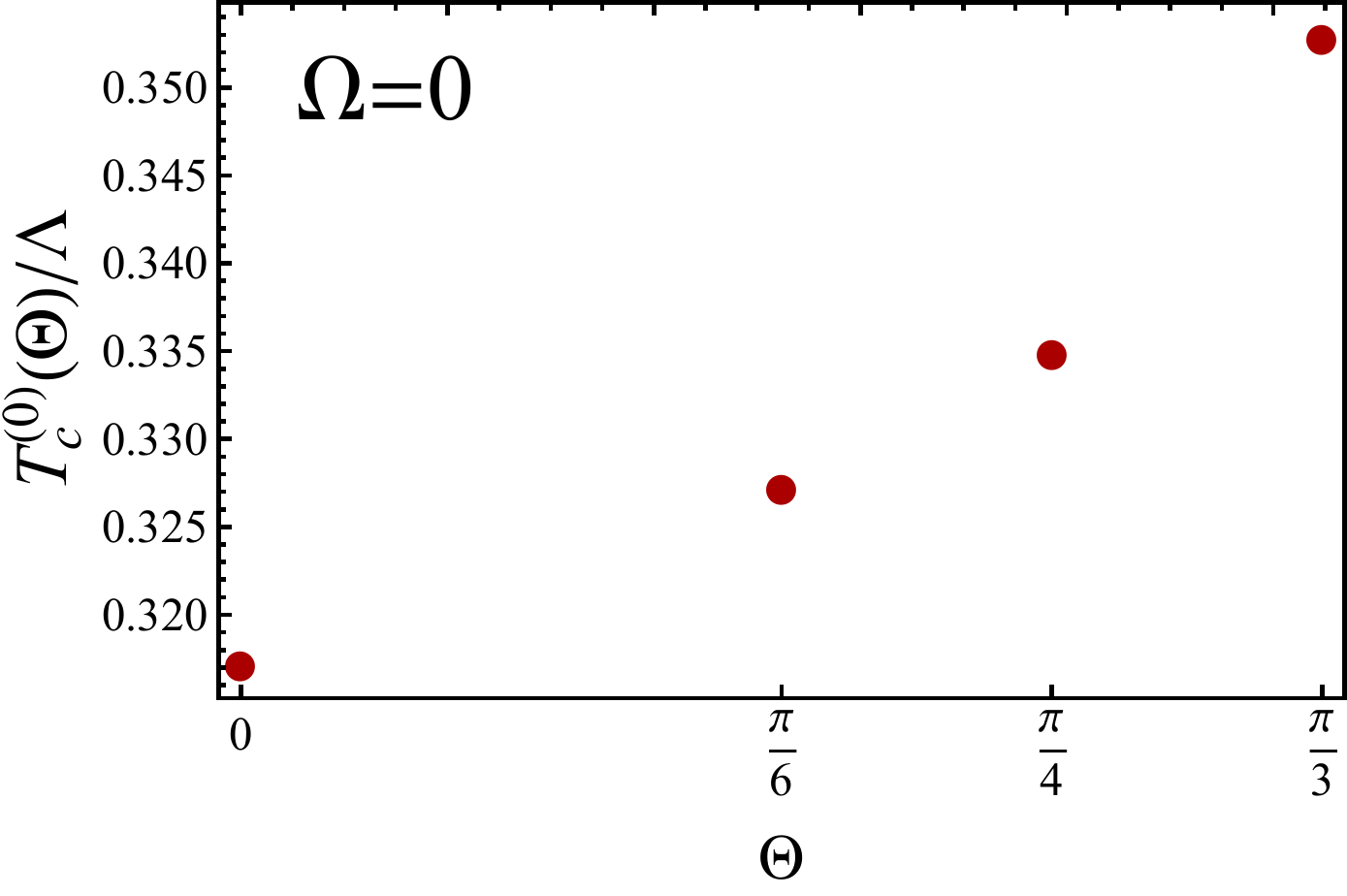}
\end{center}
\caption{Critical temperature $T_c^{(0)}$ of the chiral symmetry restoration at various angles $\Theta$ of the chiral MIT boundary condition for the radius $R = 20/\Lambda$ and the coupling $G=42/\Lambda^2$.}
\label{fig:Tc}
\end{figure}

\begin{figure}[!thb]
\begin{center}
\includegraphics[scale=0.5,clip=true]{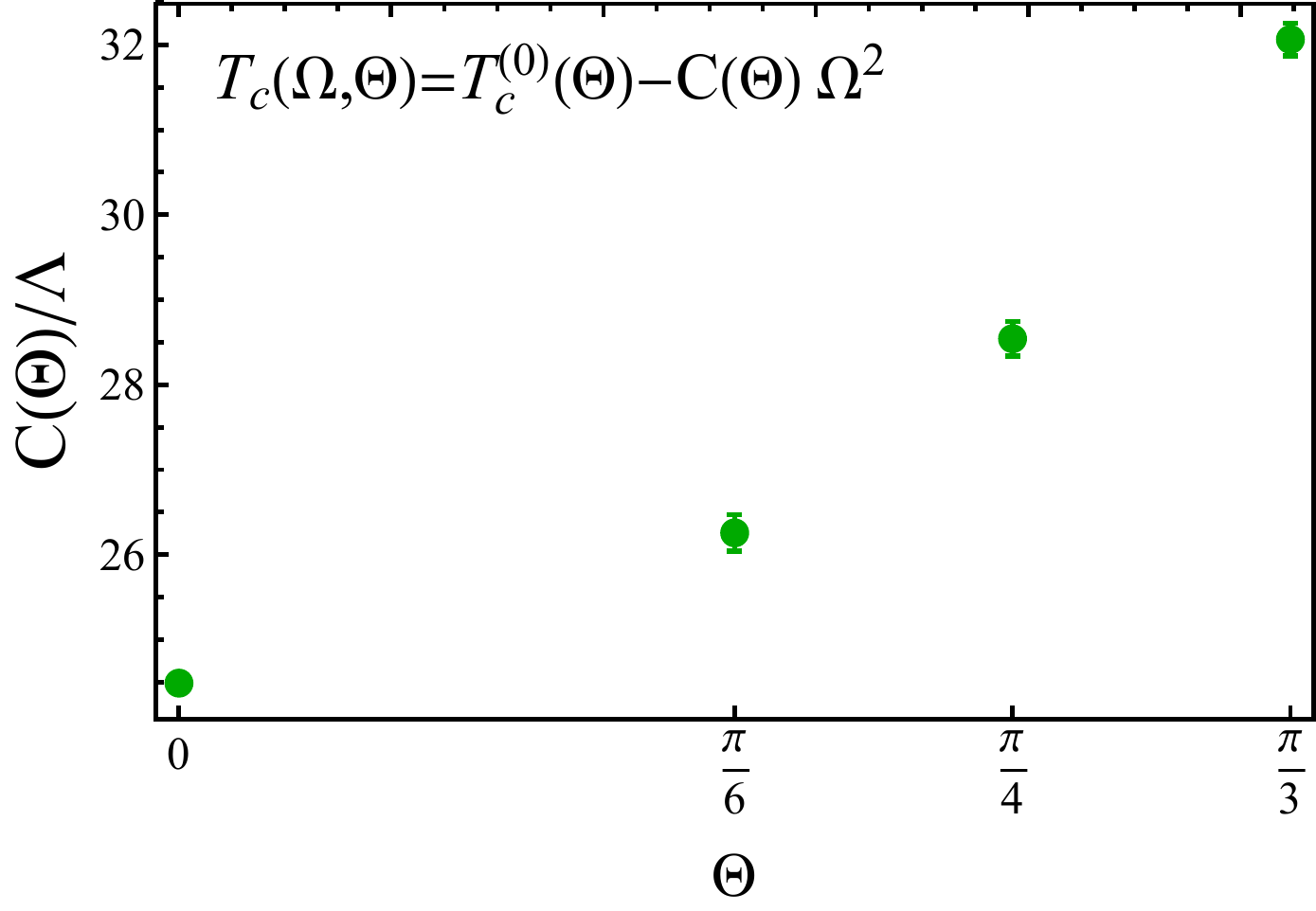}
\end{center}
\caption{The same as in Fig.~\ref{fig:Tc} but for the slope $C(\Theta)$ of the quadratic dependence of the critical temperature on the angular frequency~\eq{eq:Tc:fit}.}
\label{fig:C:Theta}
\end{figure}

The phase diagrams for all four studied boundary angles $\Theta$ are shown in Fig.~\ref{fig:phase} (the critical curve at $\Theta=0$ corresponds to our previous result~\cite{Chernodub:2016kxh}). Although all critical phase lines have a similar dependence on the angular frequency $\Omega$, the phase transitions clearly depend on the angle $\Theta$ of the chiral MIT boundary condition imposed at the cylinder which bounds the rotating fermionic matter. There are two sources of difference between the critical transition lines with different boundary angles $\Theta$:

\begin{itemize}

\item[(i)] The geometric effect of the finite radius $R$ leads to dependence of the critical temperature $T_c(\Omega=0,\Theta) \equiv T_c^{(0)}(\Theta)$ on the boundary condition $\Theta$ even in the case of a nonrotating matter. The critical temperature at $\Theta = 0$ is shown in Fig.~\ref{fig:Tc}.

\item[(ii)] The coefficient $C(\Theta)$ -- which determines the slope of the dependence~\eq{eq:Tc:fit} of the critical temperature $T_c$ on the angular frequency $\Omega$ -- is sensitive to the boundary angle $\Theta$. We have shown the slope coefficient $C(\Theta)$ in Fig.~\ref{fig:C:Theta}.

\end{itemize}

\section{Conclusions}

Any rigidly rotating system should be bounded in directions transverse to the axis of rotation in order to avoid unphysical causality-violating effects. In order to ensure that no particle propagates into the forbidden faster-than-light regions at sufficiently large distances from the axis of rotation, appropriate conditions should be imposed on particle wavefunctions at the transverse boundary of the system. Thus, any rigidly rotating system must be (i) transversally bounded and (ii) its properties should depend on boundary conditions. 

Recently there was a number of studies of dynamical mass generation phenomenon in rigidly rotating fermionic systems both in unbounded~\cite{Chen:2015hfc,Jiang:2016wvv} and bounded~\cite{Ebihara:2016fwa,Chernodub:2016kxh} geometries. In particular, general features of the phase diagram in the ``temperature - angular frequency'' plane were determined~\cite{Jiang:2016wvv,Chernodub:2016kxh}. The critical temperature of the spontaneous chiral phase transition turns out to be a diminishing function of the angular frequency $\Omega$. Notice that despite the volume of the system is infinite (as the height of the rotating cylinder along the axis of rotation is not restricted) the finite-geometry effects of the transverse directions play an important role in the dynamical symmetry breaking.

In our article we stress that the imposition of appropriate boundary conditions is an important requirement which should be satisfied by any rigidly rotating system in order to avoid unphysical pathologies~\cite{ref:Levin,Davies:1996ks,Duffy:2002ss,Ambrus:2014uqa,Ebihara:2016fwa}. This statement leads naturally to the question of how strong the dependence of the chiral phase structure of interacting and rigidly rotating fermions on the particular form of the boundary condition is. 

In order to answer the question of boundary dependence, we find the spectrum of free massive Dirac fermions confined inside a cylinder with so-called chiral MIT boundary condition~\eq{eq:BC} which is a generalization of the standard MIT boundary condition. This generalization is characterized by a continuous chiral angle $\Theta \in [0, 2\pi)$. In Ref.~\cite{Ambrus:2015lfr} the spectrum was solved for the cases $\Theta = 0$ and $\Theta = \pi$, and in our paper we generalize it to an arbitrary chiral angle $\Theta$.

Next, we determine the phase structure of interacting fermions described by the Nambu--Jona-Lasinio model at finite temperature $T$ and at fixed chiral angle $\Theta$ of the boundary condition. The phase diagrams in the $T-\Omega$ plane for different values of $\Theta$ are shown in Fig.~\ref{fig:phase}. We find that the temperature of the chiral phase transition $T_c$ depends substantially on the boundary condition of both rotating and nonrotating fermionic matter. The angle $\Theta$ of the chiral MIT boundary condition affects the rotation-independent shift of the critical temperature as well as the slope of its quadratic dependence~\eq{eq:Tc:fit} on the angular frequency $\Omega$. While the quantitative features of the phase diagram depend on boundary angle $\Theta$, its qualitative features at all studied values of $\Theta$ are the same: the critical temperature $T_c$ decreases as a quadratic function~\eq{eq:Tc:fit} of the angular frequency $\Omega$.

The same statement is also applied to the (critical) temperatures and frequencies of the discontinuities between numerous flat plateaux in the mass gap $\sigma$ in the chirally broken phase (Fig.~\ref{fig:sigma:3d}). These discontinuities represent a distant (irregular) analogue of magnetic Shubnikov--de Haas oscillations~\cite{Chernodub:2016kxh}. 

In the dynamically unbroken phase the boundary conditions lead to a small explicit chiral symmetry breaking. It is a proximity effect caused by the boundary conditions which are not invariant under the chiral transformations~\cite{Chernodub:2016kxh}. 

Thus, we demonstrated a noticeable dependence of the phase structure of the rotating interacting fermionic system on the type of the boundary conditions which should necessarily be imposed to confine fermions within a rotating volume.

\acknowledgments 

The work of S.~G. was supported by a grant from La Region Centre (France).

\end{document}